\documentclass[letterpaper, 10 pt , conference]{ieeeconf}

\IEEEoverridecommandlockouts
\usepackage[utf8]{inputenc}
\usepackage[T1]{fontenc}
\usepackage{color}
\usepackage[ruled,vlined]{algorithm2e}
\usepackage{amsfonts}
\usepackage{amsmath}
\usepackage{amssymb}
\usepackage{mathtools}
\usepackage{commath}
\usepackage{float}
\usepackage{epsfig}
\usepackage{bm}
\usepackage{lipsum}
\usepackage{cuted}

\usepackage{caption}
\usepackage{subcaption}

\newtheorem{assumption}{Assumption}

\newtheorem{proposition}{Proposition}

\begin{document}

    \title{ \LARGE \bf An Application of Model Reference Adaptive Control for Multi-Agent Synchronization in Drone Networks}

\author{{Miguel F. Arevalo-Castiblanco$^*$, Yejin Wi$^*$, Marzia Cescon and, C\'esar A. Uribe} 
\thanks{$^*$MFAC and YW are co-first authors. MFAC and CAU are with Rice University, Houston, TX, 77005, USA. (e-mail: \{mfarevalo,  cauribe\}@rice.edu). YW and MC are with the University of Houston, Houston, TX, 77204, USA. (e-mail: \{ywi2,  mcescon2\}@central.uh.edu). MFAC is with the Departamento de Ingeniería Eléctrica y Electrónica, Universidad Nacional de Colombia, Bogotá, Colombia 111321, $\{$\texttt{miarevaloc\}@unal.edu.co.} The work of CAU is partially supported by the National Science Foundation under Grants \#2211815 and \#2213568. The work of MFAC is supported by Fullbright and the IEEE Control Systems Society Graduate Collaboration Fellowship. }    }       

\maketitle
\thispagestyle{plain}
\pagestyle{plain}
    
    \begin{abstract}
This paper presents the application of a Distributed Model Reference Adaptive Control (DMRAC) strategy for robust multi-agent synchronization of a network of drones. 
The proposed approach enables the development of controllers capable of accommodating differences in real-life model parameters between agents, thereby enhancing overall network performance. We compare the performance of the adaptive control laws with classical PID controllers for the reference tracking task. Each follower drone has a model reference adaptive controller that continuously updates its parameters based on real-time feedback and reference model information. This adaptability ensures an adequate performance that, compared to conventional non-adaptive techniques, can reduce the amount of energy required and consequently increase the operating duration of the drones. The experimental results, particularly in vertical velocity control, underscore the effectiveness of the proposed approach in achieving synchronized behavior.
\end{abstract}
    
    \begin{keywords}
Networked Systems, Drones Network, Model Reference Adaptive Control, PID Control.
\end{keywords}
    
    \section{Introduction}
The proliferation of drone technology has led to diverse applications ranging from surveillance and monitoring to environmental exploration and disaster response~\cite{mcneal2016drones, wild2016exploring}. Among others, synchronization is a fundamental requirement for collaborative tasks, ensuring that drones operate harmoniously to accomplish mission objectives~\cite{santos2023development}. However, successfully deploying drone networks in real-world scenarios generates inherent challenges for efficient synchronization~\cite{alsolami2021development}. 

One key challenge in achieving synchronization in drone networks lies in variations in their dynamics and parameters, which are inevitable in the real world. These variations arise from manufacturing discrepancies, wear and tear, environmental factors, or different payloads, posing a significant challenge to developing effective control strategies for the network as a whole~\cite{cardona2021robust}. The difference between the behaviors presented by the agents in simulation compared to the behaviors that occur in reality is called the \textit{reality gap}, as given in~\cite{Tan:18}. 


This paper develops an application (from system identification and control design to physical implementation in a real drone network) of a Distributed Model Reference Adaptive Control (DMRAC) strategy designed to enable the seamless integration of adaptive control techniques into multi-agent drone systems without the use of complex methods such as projections or inverse control laws~\cite{peng2013distributed}. The primary objective of our research is to implement and evaluate the performance of an adaptive control framework that can accommodate differences in the model parameters of individual drones within a network. The DMRAC approach leverages adaptive control principles, allowing each drone to continuously update its parameters based on real-time feedback and reference model information. This adaptability ensures that the drones can autonomously and dynamically adjust their behavior, promoting synchronization in the face of varying environmental conditions and drone-specific characteristics. Furthermore, our research integrates classical control techniques into an experiment framework to validate performance by considering heterogeneous agents in the drone network without running additional identification and control design routines for each agent in the system. The framework consisted of 5 experiments: 1) an initial experiment to test MRAC and PID controllers in a centralized manner. 2) A communication experiment between drones. 3) The heterogeneous communication controller. 4) A battery-life test, and 5) the final experiment with a larger number of physical drones. This framework facilitates an examination of both the temporal response of the controllers and the energy associated with the generated control actions.

The subsequent sections of the paper are structured as follows: Section~\ref{s:quadrotor} presents the modeling and dynamic analysis of the drones used in the experimentation.  Section~\ref{s:leader} defines an identification process and vertical velocity control for the leader drone.  With the reference identified, the distributed adaptive control design is presented in Section~\ref{s:mrac}.  The validation and comparison of these technologies are presented in section~\ref{s:experiments}, to obtain conclusions and define future work in section~\ref{s:conclusions}.

\textbf{Notation.} The set of real numbers is denoted as $\mathbb{R}$. We write $\mathbf{X}^\top$,  $x^\top$, $f(\cdot)^\top$ for the transpose of a matrix, a vector, or a function, respectively.  In graph theory, a directed graph is defined as the pair $(\mathcal{V},\mathcal{E})$, where $\mathcal{V}$ is the set of graph nodes, and $\mathcal{E} \in \mathcal{V}\times \mathcal{V}$ is the set of communication edges. The adjacency matrix of the graph $\mathcal{G}$ is defined as $\mathbf{A}=[a_{ij}]$ where $a_{ii}=0$ and $a_{ij}=1$ if and only if $(j,i) \in \mathcal{E}$, where $i \neq j$.  The Laplacian matrix of a graph is defined as $\mathbf{L}=\mathbf{D}-\mathbf{A}$, with $\mathbf{D}$ as the degree matrix. Based on the $\mathbf{L}$ structure, at least one of its eigenvalues is zero, and the rest have nonnegative real parts. A digraph $\mathcal{G}$ is weight-balanced if and only if $1^\top_N\mathbf{L=0}$.


    \section{Quadrotor Dynamics and Kinematics}\label{s:quadrotor}

This section describes the hardware, including the mathematical model of the drone platform for vertical velocity control in a synchronization context. The modeling process for the control of each variable is done according to the hardware resources available for experimentation. 

\begin{figure}
    \centering
    \begin{subfigure}[b]{0.45\linewidth}
        \centering
        \includegraphics[width = \linewidth]{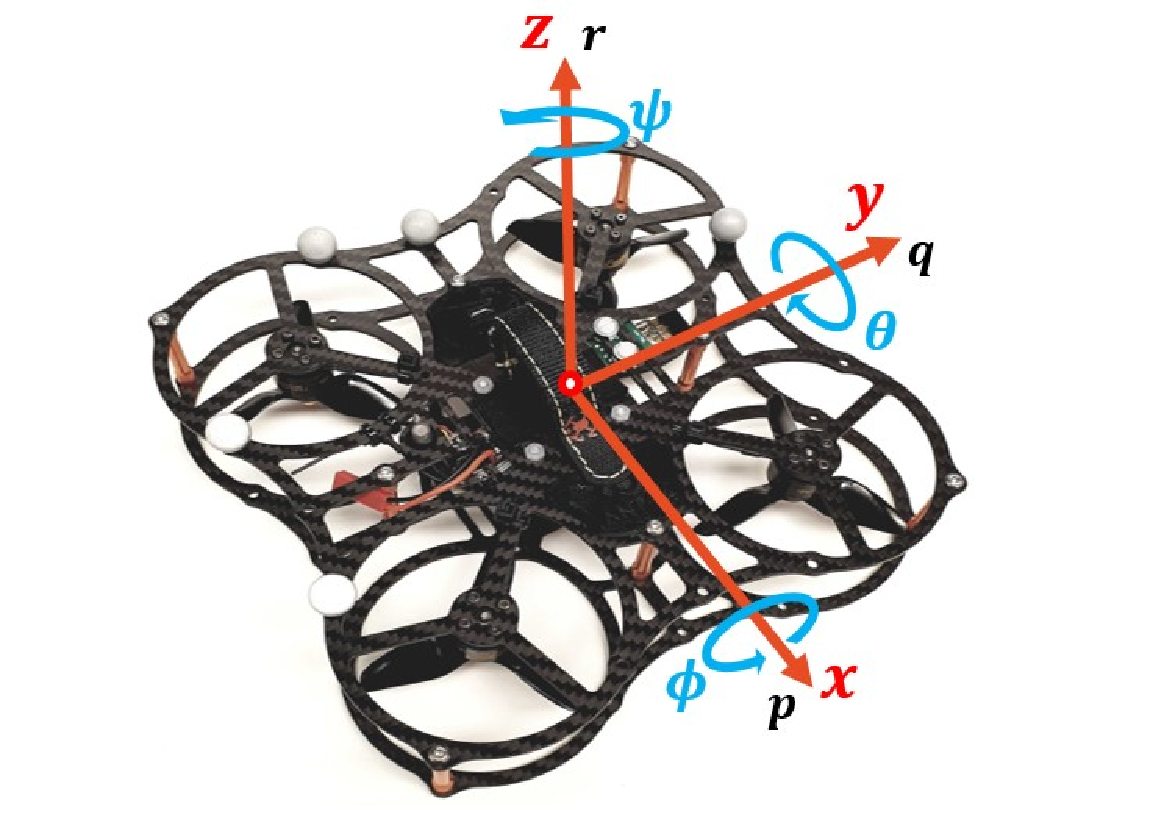}
        \caption{} 
        \label{fig:1}
    \end{subfigure}
    \begin{subfigure}[b]{0.52\linewidth}
        \centering
        \includegraphics[width = \linewidth]{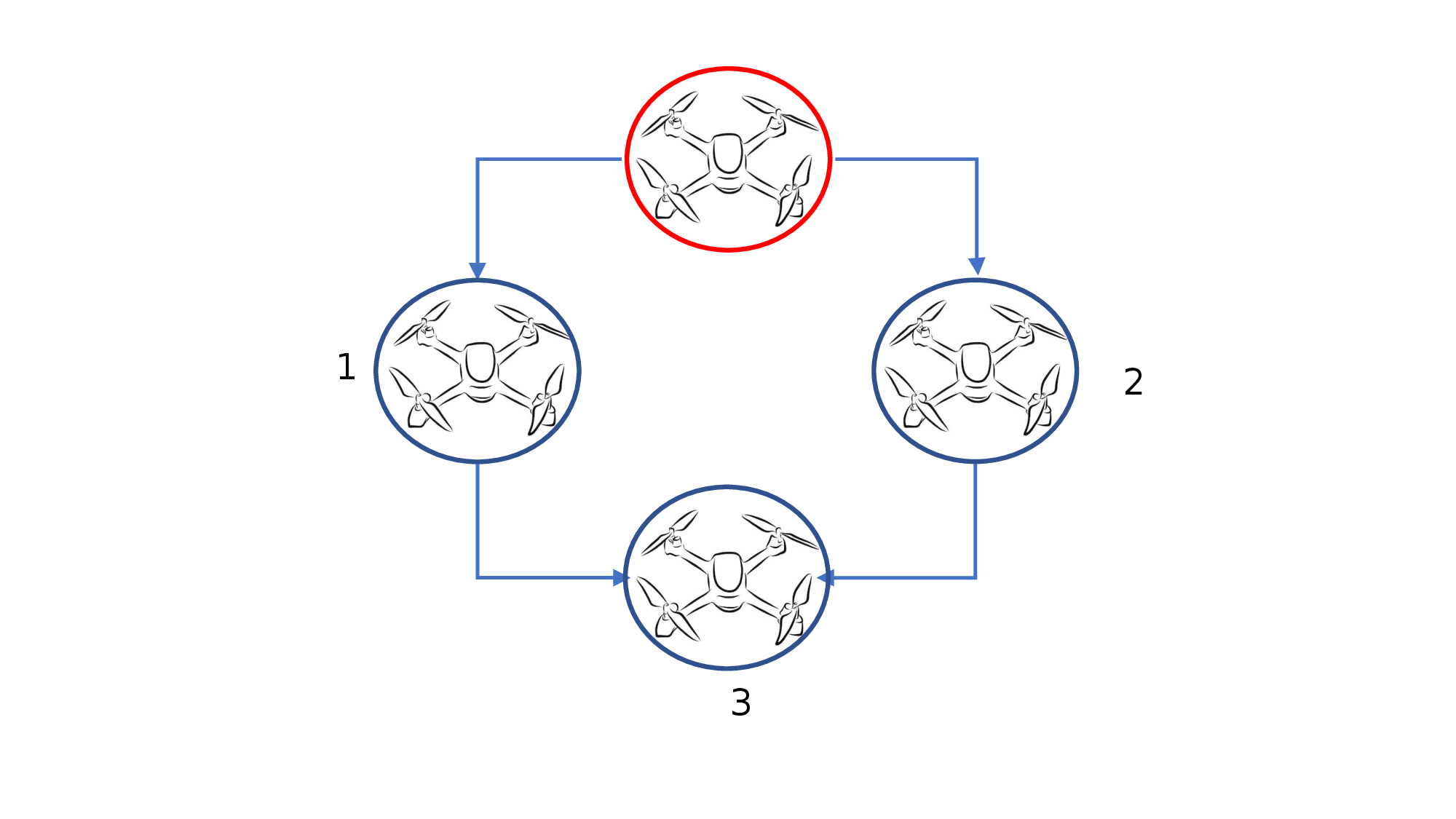}
        \caption{} 
        \label{fig:2}
    \end{subfigure}
    \caption{(a) Quadrotor body frame; angles and angular velocities. (b) Block diagram DMRAC drone implementation.} 
    \label{fig:frame_work}
\end{figure}

We consider a network of $N$ intercommunicated drones. Each drone has the configuration shown in Fig.~\ref{fig:1} composed of a quadcopter airframe with $x$, $y$, and $z$ as the body frame with red lines, the angles with black lines, and the angular velocities \textit{yaw} ($p$, $\phi$), \textit{pitch} ($q$, $\theta$) and \textit{roll} ($r$, $\psi$) with blue lines. Each drone has wireless communication with its peers within a predefined communication graph, as seen in Fig.~\ref{fig:2}. One of the drones is considered a leader (shown in red), and the rest are considered as followers (shown in blue). All the agents have a set of dynamic variables. We denoted $x_m$ to the leader set of variables, and $x_i\; \forall\in\mathcal{V}$ for the follower agents. 

Synchronization is measured through the difference in the state variables of two agents. Synchronization errors can be denoted as $e_i=x_m-x_i$ or $e_{ij}=x_i-x_j$.

Given a desired trajectory for a drone network to follow, with a desired position, velocity, and acceleration defined as $\mathbf{r}_{b}^d$, $\mathbf{v}_{b}^d$, and $\mathbf{\dot{v}}_{b}^d$, respectively, the objective is to design a control input$f_i,\boldsymbol{\tau}_i$ for each quadrotor so that the synchronization error tends to zero through wireless communication.

\subsection{System Architecture}

For all the experimentation process, we used the ANT-X drone, described with the body-fixed frame in Fig.~\ref{fig:1}, a quadcopter designed for academic and research purposes. General specifications of the vendor can be found at \textit{https://antx.it/educational-products-antx/dronelab-antx/}. 

\begin{figure}[t]
\centering
\includegraphics[width = 0.48\textwidth]{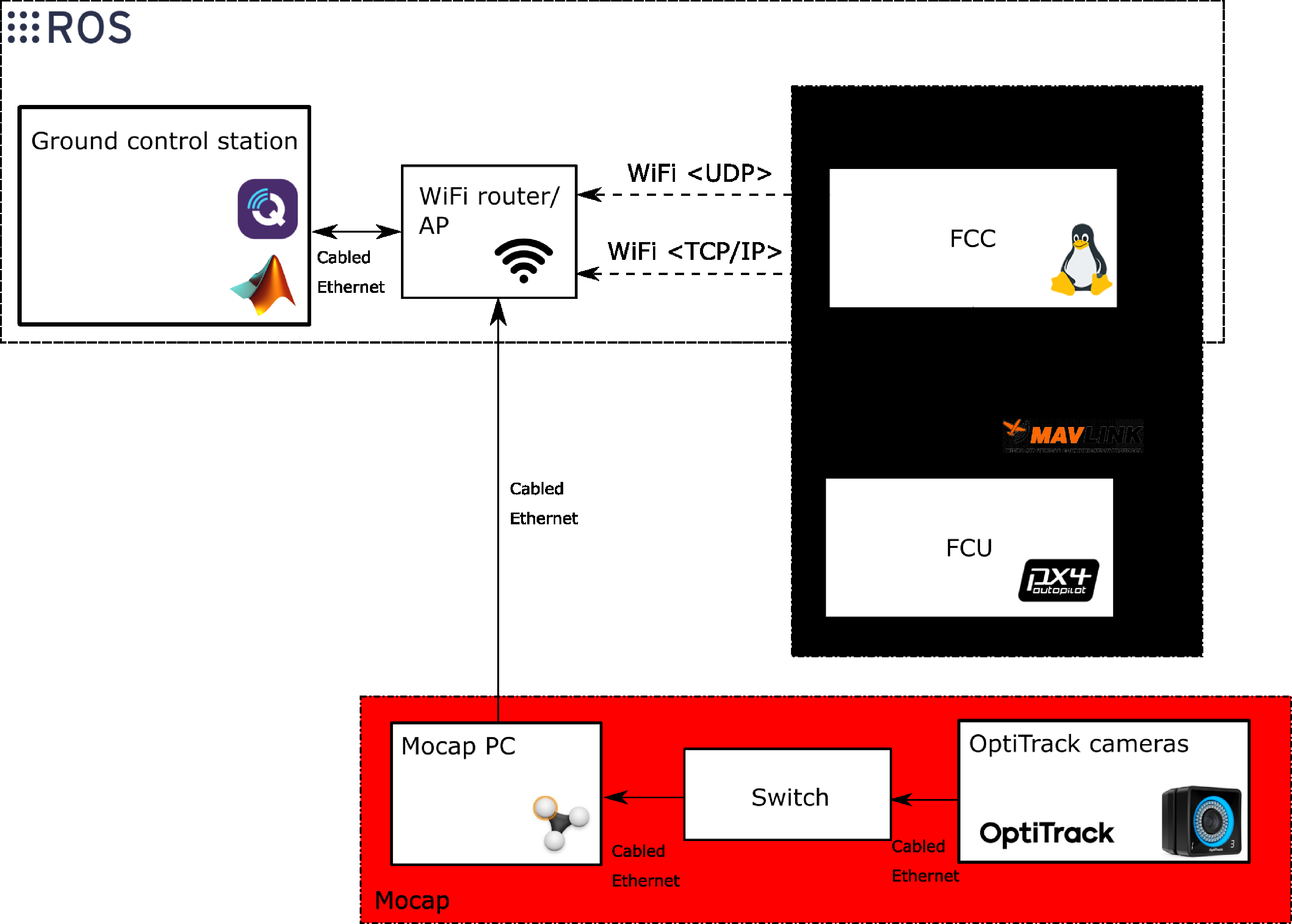}
\caption{Software architecture for implementation in ANT-X drones.}	
\label{fig:software}
\end{figure}

The ANT-X is a quadcopter with six degrees of freedom (6DOF), operated using its own hardware and software systems as presented in Fig.~\ref{fig:software}. The hardware system comprises a lightweight quadcopter, a Flight Control Unit (FCU) equipped with an Inertial Measurement Unit (IMU) to measure the parameters of the Unmanned Aerial Vehicle (UAV), a Flight Companion Computer (FCC) for transmission, a Ground Control Station (GCS) for interacting with users, and OptiTrack cameras to obtain the quadcopter's position in a flight volume. The quadcopter's software system consists of MATLAB, Simulink, PX4 Autopilot, an open-source flight control program for drones, Robot Operating Systems (ROS), which facilitate the implementation of the system, and motion capture system (MoCap) for transmission of the captured drone's information via OptiTrack cameras. 

Each drone is equipped with FCC running a Linux distribution that provides high-level computing capabilities and allows communication with the GCS via WiFi. Additionally, the user can customize them using the SLXtoPX4 software tool, which allows high-level controllers to be implemented in Simulink, automatically generate the controller code, and integrate it with the firmware. 

\subsection{Dynamic Model}

The model describing the attitude dynamics is given by the conservation of angular momentum defined in the quadrotor's body frame as:
\begin{equation}
    J\dot{\Omega}+\Omega\times J\Omega=T_c+T_e,
\end{equation}
where we denote the inertia matrix referred to the center of mass of the quadrotor with $J \in \mathbb{R}^{3 \times 3}$, the vector of the angular velocities about the $x-$, $y-$ and $z-$ axis, respectively, in the quadrotor body frame depicted in Fig.~\ref{fig:1} with $\Omega=[p \; q\; r]^\top\in\mathbb{R}^3$,  $T_c=[L_c \; M_c \; N_c]^\top\in\mathbb{R}^3$ is the control torque acting at the center of mass of the quadrotor delivered by the propellers, $T_e=[L_e \; M_e \; N_e]^\top\in\mathbb{R}^3$ is the exogenous torque acting at the center of mass of the quadrotor collecting the effect of exogenous forces and torques (gravity, aerodynamics, and friction), $\phi,\theta,\psi$ are the \textit{yaw}, \textit{pitch}, \textit{roll} angles respectively.

The attitude kinematics describing the time evolution of \textit{yaw}, \textit{pitch}, and \textit{roll} angles due to angular velocity is:
\begin{equation}
    \begin{bmatrix}
\dot{\phi} \\
\dot{\theta} \\
\dot{\psi}
\end{bmatrix}=\begin{bmatrix}
    1 & -\sin (\phi) \tan (\theta) & \cos (\theta)\tan (\theta) \\
    0 & \cos (\phi) & -\sin (\phi) \\
    0 & \frac{\sin (\phi)}{\cos (\phi)} & \frac{\cos (\phi)}{\cos (\theta)},
\end{bmatrix}\begin{bmatrix}
        p \\
        q \\
        r
    \end{bmatrix}.
\end{equation}
Further, the model describing translational dynamics and kinematics is described by:
\begin{subequations}
\begin{align}
    \dot{S}&=V, \\
    m\dot{V}&=mgE_3-T_{tot}B_3(\phi,\theta,\psi)+F_e,
\end{align}
\end{subequations}
where $\
S=[x \; y \; z]^\top$ is the position vector of the quadrotor center of mass for the original vector frame; $V=[v_x \; v_y \; v_z]^\top$ is the velocity of the quadrotor center of mass resolved in the inertial frame; $mgE_3=mg[0 \; 0 \; 1]^\top$ is the force associated with gravity determined in the inertial frame; $F_e=[f_{xe} \; f_{ye} \; f_{ze}]^\top$ is a disturbance force vector and, $T_{tot}$ is the total thrust delivered by the propellers, which is directed along the negative direction of the vertical axis of the quadrotor frame, represented in the inertial frame by the unit vector:
\begin{equation}
    B_3(\phi,\theta,\psi)=
    \begin{bmatrix}
        \sin(\phi)\sin(\psi)+\cos(\phi)\sin(\theta)\cos(\psi) \\
        -\sin(\phi)\cos(\psi)+\cos(\phi)\sin(\theta)\sin(\psi) \\
        \cos(\phi)\cos(\theta)
    \end{bmatrix}.
\end{equation}
Linearized attitude and translational dynamics are obtained by linearizing the equations of motion concerning the hovering condition, which is an equilibrium point characterized by low velocities and slight deviations from a fixed position. In particular, for the case under consideration in this work, the linearized vertical dynamics ($z-$ axis) with $\Delta T_{tot}=T_{tot}-mg$ is expressed as:
\begin{subequations}
\begin{align}\label{eq:linear_vertical}
    \Delta\dot{ z}&=\Delta v_z, \\
    m\Delta\dot{ v_z}&=-\Delta T_{tot} + f_{ze},
    \label{eq:linear_vertical1}
\end{align}
\end{subequations}
where $f_{ze}$ is the vertical component of the disturbance vector.
Defining $F_z=-T_{tot}$, with $F_z$ [N] vertical force, ~\eqref{eq:linear_vertical1} becomes
\begin{equation}
m \Delta \dot{v_z}=\Delta F_z + f_{ze},
\end{equation}
where $\Delta F_z=F_z+mg$.
Taking Laplace transform, the linearized model can be conveniently written as:
\begin{equation}
    v_z(s) = G_{vz}(s) F_z(s),
    \label{eq:9}
\end{equation}
with $G_{vz}$ the transfer function from vertical force to vertical velocity.

Based on the dynamic expressions obtained for the vertical velocity model $G_{vz}$ in \eqref{eq:9}, the next section shows the system identification from in-flight experimental data to parameterize and control the leader drone in the experiments. 

    \section{Leader drone parameterization}\label{s:leader}
This section describes the parameterization used to control the lead drone. An identification process is initially developed to obtain the lead agent model parameters. With the identified dynamics, the design of a PID controller is shown.

\subsection{System Identification}
\begin{figure}
    \centering
    \begin{subfigure}[b]{0.45\linewidth}
        \centering
        \includegraphics[width = \linewidth]{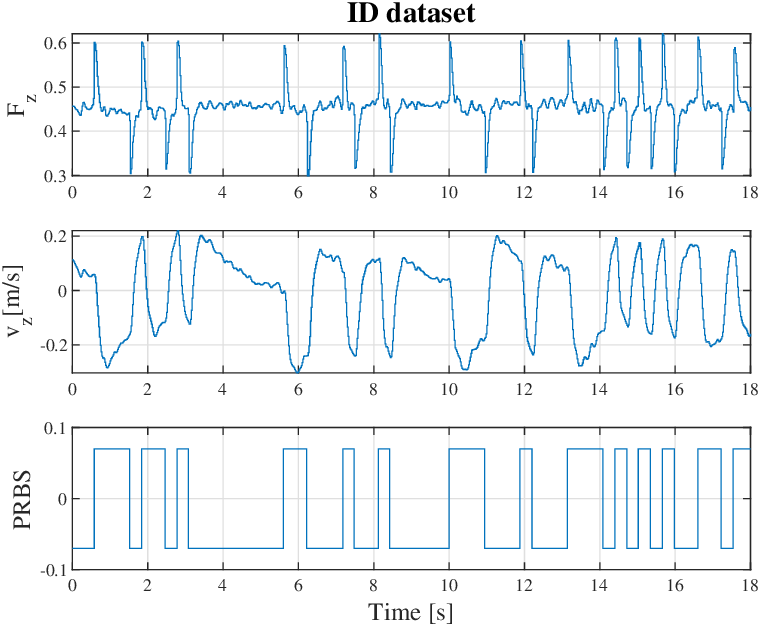}
        \caption{}
        \label{fig:sysIDdata}
    \end{subfigure}
    \begin{subfigure}[b]{0.45\linewidth}
        \centering
        \includegraphics[width = \linewidth]{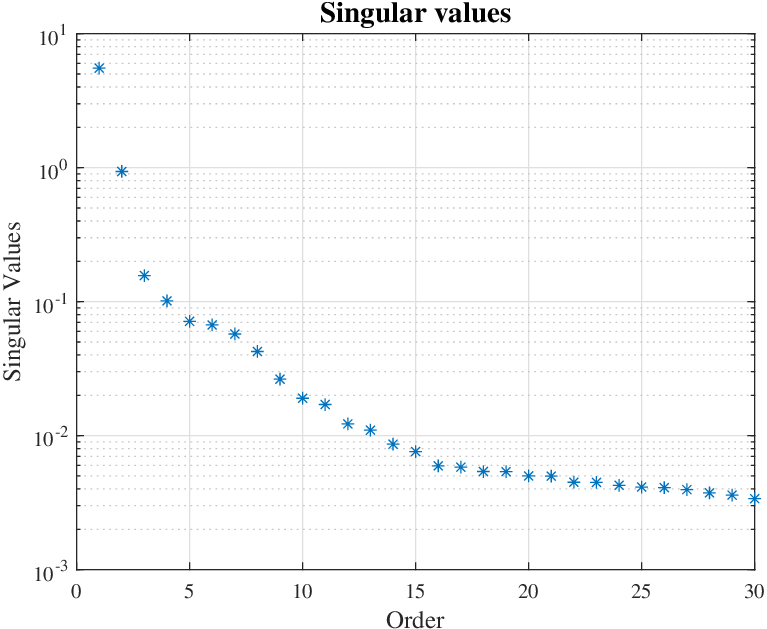}
        \caption{} 
        \label{fig:singular}
    \end{subfigure}
    \caption{(a) Identification data. (b) Singular values.} 
    \label{fig:systemID}
\end{figure}

Identification experiments were conducted under closed-loop position and attitude control to excite the vertical dynamics in the frequency range 0-10 rad/sec, which is the expected bandwidth of the vertical velocity control, by injecting a Pseudorandom Binary Sequence (PRBS) input signal into the vertical axis. Fig.~\ref{fig:sysIDdata} illustrates the collected data. $F_z$ and $v_z$ are the input and outputs, force and velocity, respectively, and PRBS is the injected excitation signal to the input, $v_z$. An equivalent time delay due to computation and communication lags was estimated at $\tau_e=0.02$ sec by visual inspection of the data. The effect of $\tau_e$ was removed by shifting the output backward for accurate modeling. The Predictor-Based Subspace Identification method \texttt{PBSID}$_{\text{opt}}$~\cite{van2013closed} was used to identify a discrete-time, linear time-invariant (LTI) state space model with past and future horizons, $p$ and $f$ respectively, set to be $p = f = 30$, and model order $n = 2$, selected after inspection of the singular values in Fig.~\ref{fig:singular}.
The identified state-space system matrices are obtained as:
\begin{subequations}
\begin{align*}\label{state-space}
    A&=\begin{bmatrix}
        1.01 & 0.15 \\
        -0.02 & 0.74
    \end{bmatrix}, &B=\begin{bmatrix}
        0.13 \\
        0.22
    \end{bmatrix}, \\
    C&=\begin{bmatrix}
        -1.84 & 0.31
    \end{bmatrix}, &D=0,
\end{align*}
\end{subequations}
with corresponding transfer function expressed in discrete and continuous time as
\begin{equation}
    G_{vz}(z)=z^{-1}\left(\frac{-0.0013z^2+0.017z-0.005}{z^2-1.75z+0.75}\right),
\end{equation}
\begin{equation}
    G_{vz}(s)=e^{-0.02s}\left(\frac{0.319s+29.92}{s^2+14.03s+3.099}\right).
\end{equation}

\begin{figure}
    \centering
    \begin{subfigure}[b]{0.45\linewidth}
        \centering
        \includegraphics[width = \linewidth]{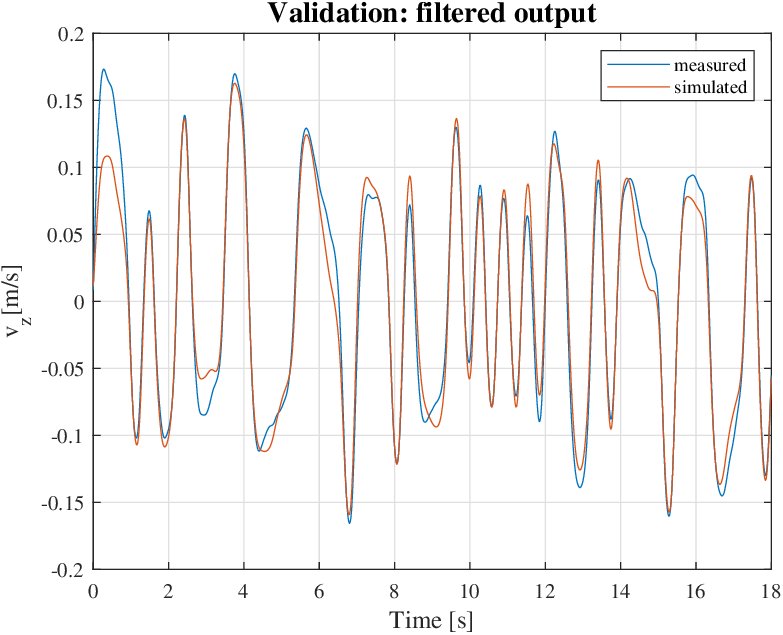}
        \caption{}
        \label{fig:time_domain}
        \end{subfigure}
    \begin{subfigure}[b]{0.45\linewidth}
        \centering
        \includegraphics[width = \linewidth]{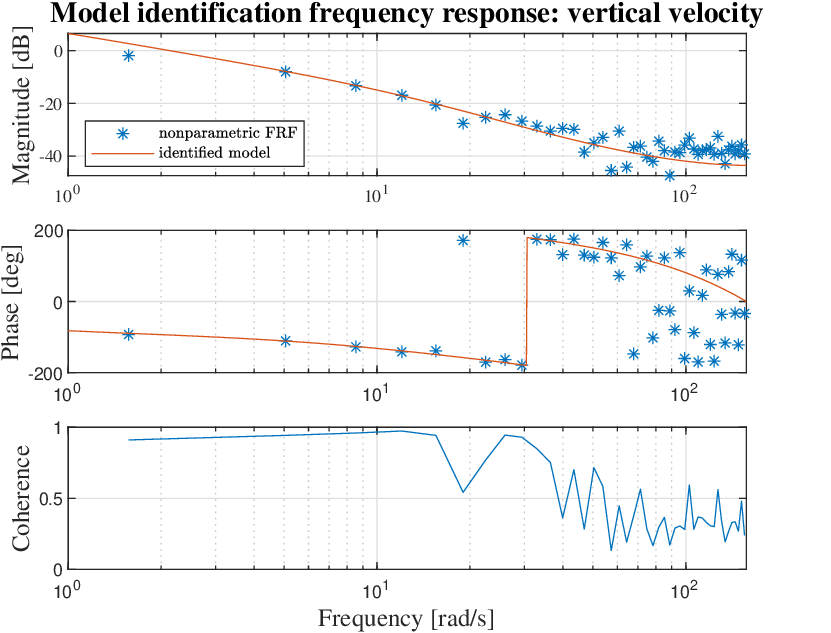}
        \caption{}
        \label{fig:frequency_domain}
        \end{subfigure}
    \caption{(a) Cross-validation. (b) Verification of identified vertical dynamic.} 
    \label{fig:systemID_1}
\end{figure}

The accuracy of the identified model in the frequency range of interest was verified by comparing its Bode diagram against a nonparametric frequency response. In Fig. ~\ref{fig:frequency_domain}, the identified model's frequency response matches the measured model's frequency response in the 2-20 rad/s range. 
Cross-validation using a completely separate dataset revealed a Variance Accounted For (VAF) of 96\%. Figure~\ref{fig:time_domain} gives a time-domain representation of validation results.


\subsection{Baseline PID Controller}

\begin{figure}
\centering
\includegraphics[width = .5\textwidth]{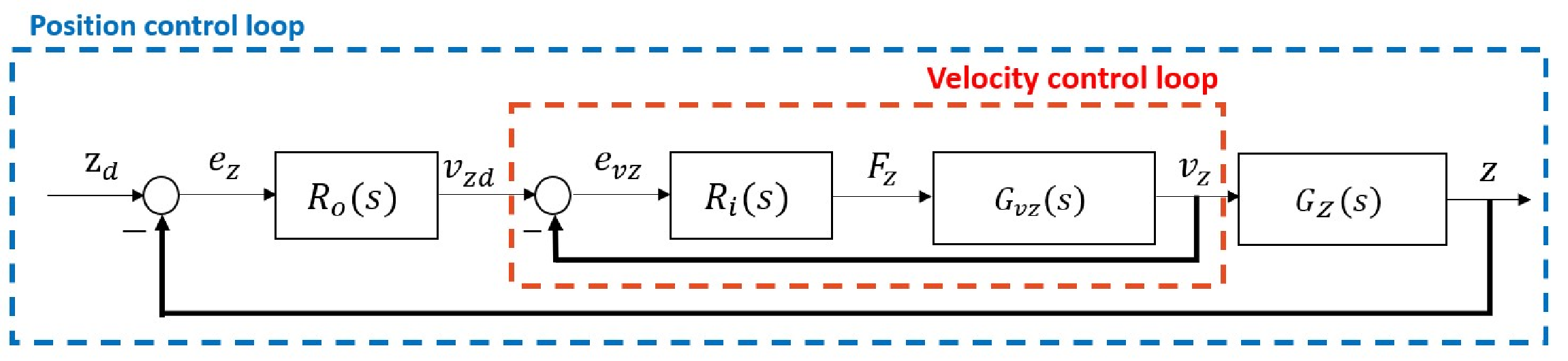}
\caption{Block diagram for altitude control.}	
\label{fig:block_diagram}
\end{figure}

This subsection describes the classical PID controller designed for the reference quadrotor. Graphically, Fig.~\ref{fig:block_diagram} shows the cascaded architecture implemented for vertical position and velocity control. According to the scheme, the outer loop computes a reference vertical velocity $v_{zd}$ from the vertical position error $e_z = z_d - z$. In contrast, the inner loop computes the vertical force $F_z$ required to track the reference angular velocity $v_{zd}$. We designed the inner loop controller $R_i(s)$ as a classical PID. 

\begin{figure}[t]
\centering
\includegraphics[width = 0.9\linewidth]{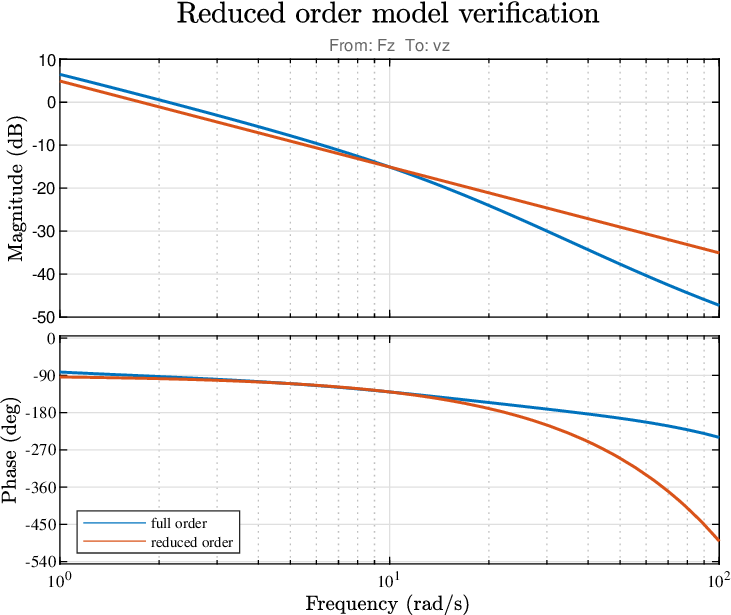}
\caption{Verification of simplified model.}	
\label{fig:reduction}
\end{figure}

According to \cite{papadopoulos1988pole}, Model Order Reduction (MOR) is crucial to investigating and simulating the actual model and decreases the intricate computation in the phase of controller development. The MOR produces a simplified model with a specific matched frequency to the original model; the significantly relevant characteristics to the original system are maintained in the reduced order model \cite{panza2015rotor}. The following conditions are considered:

\begin{itemize}
\item The magnitude of the original model at a certain frequency \begin{math}\omega_1\end{math}, is identical to the magnitude of the reduced order model at the same frequency  \begin{math}\omega_1\end{math}:
\begin{equation}\label{eq:19}
\mid{G_{vz,red}}(j\omega_1)\mid = \mid{G_{vz}}(j\omega_1)\mid.
\end{equation}

\item The phase of the original model at a certain frequency \begin{math}\omega_2\end{math}, is identical to the phase of the reduced order model at the same frequency \begin{math}\omega_2\end{math}:
\begin{equation}\label{eq:20}
\angle({G_{vz,red}}(j\omega_2)) = \angle({G_{vz}}(j\omega_2)).
\end{equation}
\end{itemize}
The reduced order model is ascertained by comparing it with the frequency response of the original model in Fig.~\ref{fig:reduction}. The frequency responses of the full-order and simplified models are indistinguishable at the frequency of 10 rad/s; the main characteristic of the original models remains. 
 
 The simplified model is applied to obtain the classical PID controller, which is formulated as
\begin{subequations} \label{eq:PID}
\begin{align} 
    R_i(s) &= k_p + \frac{k_i}{s} + k_d\frac{s}{\tau_fs+1}, \\
        &\simeq k_i\frac{(1+\tau_1s)(1+\tau_2s)}{s(\tau_fs+1)},  
\end{align}
\end{subequations}
where \begin{math}k_p\end{math}, \begin{math}k_i\end{math} and \begin{math}k_d\end{math} are proportional, integral and derivative gain, respectively, \begin{math}\tau_f\end{math} is a derivative filter constant. In contrast, the two zeros, \begin{math}\tau_1\end{math} and \begin{math}\tau_2\end{math}, satisfy the two relationships \begin{math}\frac{k_p + k_i\tau_f}{k_i} = \tau_1 + \tau_2\end{math} and \begin{math}\frac{k_p\tau_f + k_d}{k_i} = \tau_1\tau_2\end{math}.

 The velocity control system's open-loop is shaped to obtain the PID controller with corresponding requirements based on the open-loop transfer function's desired crossover frequency, \begin{math}\omega^d_c\end{math}.
The conditions of loop shaping for the inner loop controller are introduced in the following:

\begin{itemize}
\item Place one zero, \begin{math}\tau_1\end{math}, at low frequency and another zero,\begin{math}\tau_2\end{math}, over \begin{math}\omega^d_c\end{math} to cross the 0dB axis with the slope of -20\textit{dB/dec}: \begin{math}\tau_1 = \frac{1}{\omega^d_c*\textit{a}}\end{math} and \begin{math}\tau_2 = \frac{1}{\textit{b}\omega^d_c}\end{math}. The hyperparameters, \textit{a} and \textit{b}, are chosen as 0.1 and 4 for the vertical velocity controller.

\item Derivative filter pole, \begin{math}\tau_f\end{math}, should be set at appropriately high frequency regarding \begin{math}\omega^d_c\end{math}.

\item The gain, \begin{math}\mu\end{math}, is computed through \begin{math} \mu = \omega^d_c\frac{1}{\tau_1}\end{math} to make the system crosses the 0dB axis at \begin{math}\omega^d_c\end{math}, in terms of the inner open-loop transfer function's magnitude, \begin{math}\mid{L_i}(j\omega^d_c)\mid\approx\frac{\mu\tau_1}{\omega^d_c}\end{math}.
\end{itemize}
 The revised open-loop transfer function is in the form:
\begin{equation}\
L_{red} = R_iG_{vz,red} = k^i_i\frac{\mu_{red}}{s^2}\frac{(\tau_1s + 1)(\tau_2s + 1)}{\tau_fs + 1}e^{-s\tau_{red}},
\end{equation}
where $\mu_{red}$ is the gain obtained through loop shaping according to the demands. 
The computed parameters are applied to tune the PID gains as follows:
\begin{subequations}\label{eq:28}
\begin{align}
    k_p &= (\tau_1 + \tau_2)k_i - k_i\tau_f, \\ 
    k_i &= \frac{\mu}{\mu_{red}}, \\ 
    k_d &= \tau_1\tau_2k_i - k_p\tau_f.
\end{align}
\end{subequations}
The desired crossover frequency, \begin{math}\omega^d_c\end{math}, is manually selected to 5 rad/s, and derivative filter pole's locations, \begin{math}\tau_f\end{math}, are selected to 0.022 for vertical velocity control. The PID controller developed is used for the reference drone in the network. The following section presents the distributed model reference adaptive control for the followers.

    \section{Followers Adaptive Control Design}\label{s:mrac}

Each of the follower agents $i \in\left[1,\ldots,N\right]$ is represented as a linear system  in the form
\begin{equation}
    \Dot{x}_i=A_ix_i+B_iu_i, \hspace{0.5cm} i \in \left[1,...,N\right],
    \label{eq1}
\end{equation}
where $x_i$ $\in$ $\mathbb{R}^n$ is the agent's states, $u_i$ $\in$ $\mathbb{R}^p$ is the input, $A_i$ is an unknown matrix related to the drone's states, $B_i$ are known vectors with possibly heterogeneous agents ($A_i{\not=}A_j$ and $B_i{\not=}B_j$).

In a leader-follower synchronization context, defining the dynamics of the reference pattern to be followed by the entire network is important. The reference model dynamics are described as
\begin{equation}
    \Dot{x}_m={A_m}x_m+{B_m}r,
    \label{eq2}
\end{equation}
where $x_m$ $\in$ $\mathbb{R}^n$ is the state, $r$ $\in$ $\mathbb{R}^p$ is the reference, and $A_m$ and $B_m$ are the dynamic matrices. Moreover, we assume all drones communicate over a network $\mathcal{G}(\mathcal{V},\mathcal{E})$, with $\mathcal{V}= [1,\cdots,N]$ the set of drones, and $\mathcal{E}$ the set of communication edges, such that $(j,i)\in \mathcal{E}$ if drone $j$ is an in-neighbor of drone $i$. 

\begin{assumption}\label{assum:graphs}
The communications graph $\mathcal{G}$ is unweighted, directed, and acyclic. The graph contains a directed spanning tree with the leader drone as the root node.
\end{assumption}

Assumption~\ref{assum:graphs} delimits the characteristics of the communication graph necessary for the DMRAC. Each drone can communicate the states and control actions to their in-neighbors on the network. The states and input information are used for the follower drone to generate a control action that synchronizes its dynamics with the leader by communicating with it. 
However, the discrepancy between the leader drone parameters $(A_m,B_m)$ and the follower drone parameters $(A_i,B_i)$ would make it difficult to use the same controller for the entire network. Thus, we seek for them to synchronize their states with their neighbors. 

\begin{assumption}\label{assum:feedback-mc}
For all $i \in \mathcal{V}$ there exists a vector $k_{mi}^*$ and a scalar $k_{ri}^*$ such that
\begin{equation}
    A_i+\lambda B_mk^{*\top}_{mi}=A_m \; \text{ , }\lambda k^*_{ri}B_m=B_m,
    \label{fmc}
\end{equation}
whit $\lambda>0$ an unknown scalar. Constants $k_{mi}^*$ and $k_{ri}^*$ in \eqref{fmc} are known as feedback matching conditions.
\end{assumption}

\begin{assumption}\label{assum:coupling-mc}
For all $i \in \mathcal{V}$ there exists a vector $k_{ij}^*$ and a scalar $k_{rij}^*$ such that
\begin{equation}
    A_j=A_i+{B_i}k_{ij}^{*\top}\; \text{ , }
    B_i=\lambda{B_j}k^*_{rij},
    \label{MC}
\end{equation}
where constants in \eqref{MC} are known as coupling matching conditions. 
\end{assumption}

Assumption~\ref{assum:feedback-mc} and Assumption~\ref{assum:coupling-mc} allow the matching of the dynamics of a drone to a leader and the neighbors through appropriate gains~\cite{Baldi2019}. For agents that communicate directly with the leader, the control law is defined as
\begin{equation}
    u_1=k_{m}x_m+k_{r}r,
    \label{eq6a}
\end{equation} 
where $k_{m}\in\mathbb{R}^p\times\mathbb{R}^n$ is the constant associated with the reference states, $k_{r}\in\mathbb{R}^p\times\mathbb{R}^q$ is associated with the reference signal and their adaptive laws are given by
\begin{subequations}\label{k1s}
\begin{align}
    {\dot{k}^\top_{m}}&=-{\gamma}\: {B^\top_m}P\left(x_1-x_m\right)x_1^\top,\\
    {\dot{k}_{r}}&=-{\gamma}\: {B^\top_m}P\left(x_1-x_m\right)r.
\end{align}
\end{subequations}

An adaptive gain $\gamma>0$ is defined, and it is related to the speed of adaptation of the law, for dynamics with quadcopters, the values usually are not so high to avoid oscillations in the adaptation~\cite{ntousakis2015microscopic}. $P$ can be obtained by
\begin{equation}
    PA_m+{A^\top_m}P=-Q, \hspace{0.5cm}Q\succ 0.
    \label{eq8}
\end{equation}

The following proposition presents the model reference adaptive control base case~\cite{Nguyen2008}.

\begin{proposition}[Chapter 5 in~\cite{Nguyen2018L}]
    Let Assumption~\ref{assum:feedback-mc} hold, and consider a drone $x_1$ with dynamics \eqref{eq1} communicates with a leader drone $x_m$ with dynamics~\eqref{eq2}. Then, the control law~\eqref{eq6a} with the adaptive laws~\eqref{k1s} guarantees that the synchronization error $e_1=x_1-x_m$ is bounded.
\end{proposition}

Likewise, for agents that do not have direct communication with the reference, the following control law is defined
\begin{equation}
    u_2=k^{\top}_{21}x_1+{k^\top_{m2}}\left(x_2-x_1\right)+k_{r21}u_1,
    \label{eq10}
\end{equation}
with the adaptive laws
\begin{subequations}\label{eq11}
\begin{align}
     {\dot{k}_{21}}^{\top}&=-{\gamma}\: {B^{\top}_m}P\left(x_2-x_1\right)x_2^\top,\\
    {\dot{k}_{m2}}^{\top}&=-{\gamma}\: {B^{\top}_m}P\left(x_2-x_1\right)\left(x_2-x_1\right)^\top,\\
    {\dot{k}_{r21}}&=-{\gamma}\: {B^{\top}_m}P\left(x_2-x_1\right)u_1,
\end{align}
\end{subequations}

From this approach, the following proposition is defined.

\begin{proposition}[Theorem 1 in \cite{Baldi2019}]
    Let Assumptions~\ref{assum:graphs}, \ref{assum:feedback-mc}, and~\ref{assum:coupling-mc} hold. Consider a second follower drone with dynamics~\eqref{eq1}, which is not directly connected to the leader with dynamics~\eqref{eq2}, then with the controller~\eqref{eq10} and the adaptive laws~\eqref{eq11} the synchronization error $e_{21}=x_2-x_1$ is bounded.
\end{proposition}

The experimentation framework is executed with the controllers developed in the following section.
    
    \section{Comparative Performance of PID and MRAC Controllers} \label{s:experiments}

\begin{figure}[t]
\centering
\includegraphics[width = 0.9\linewidth]{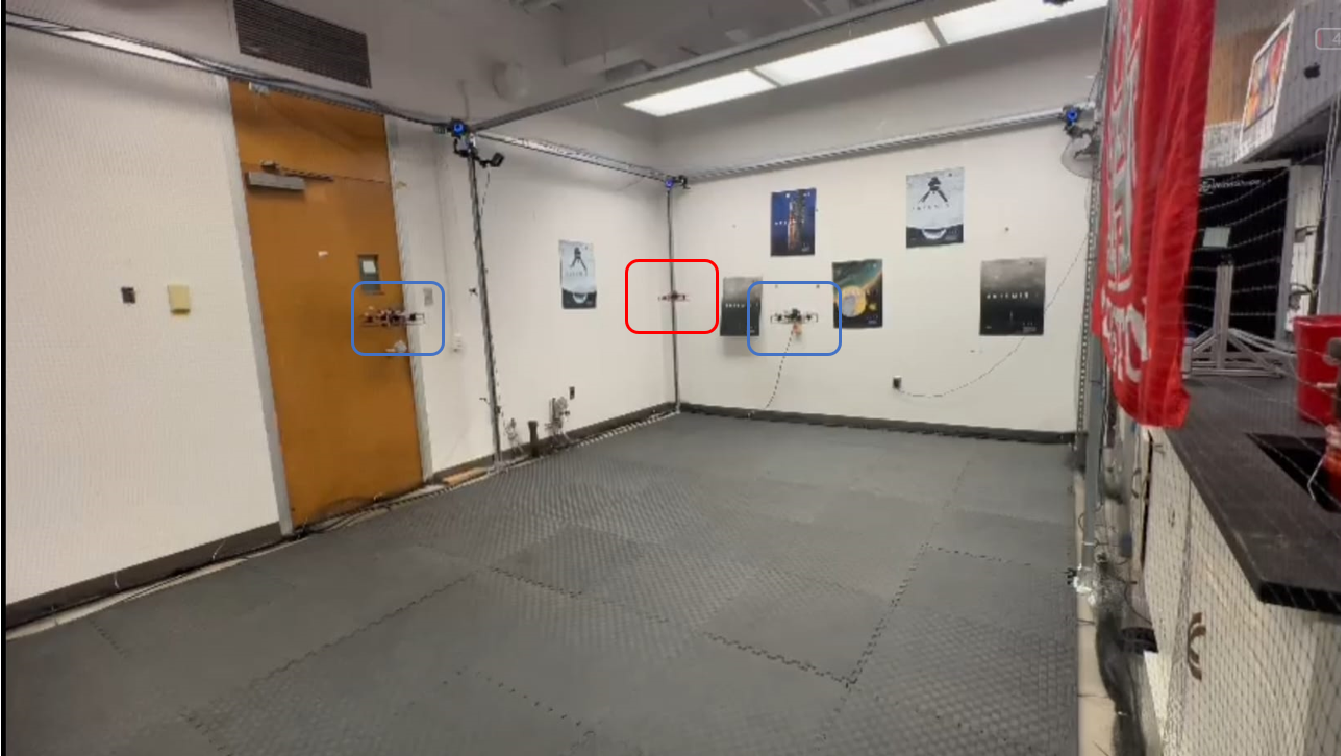}
\caption{Setup for drone network experimentation.}	
\label{fig:setup}
\end{figure}

This section presents the communication and synchronization experiments with the designed PID and MRAC controllers for vertical velocity trajectory tracking. Fig.~\ref{fig:setup} displays the physical experiment setup, which indicates one reference drone in the red box and two follower drones in the blue box. 

We propose five experiments executed under different experimental settings to observe and compare both controllers' performance regarding robustness, the change of physical parameters, and network communication. 

\begin{figure}
    \centering
    \begin{subfigure}[b]{0.49\linewidth}
        \centering
        \includegraphics[width = \linewidth]{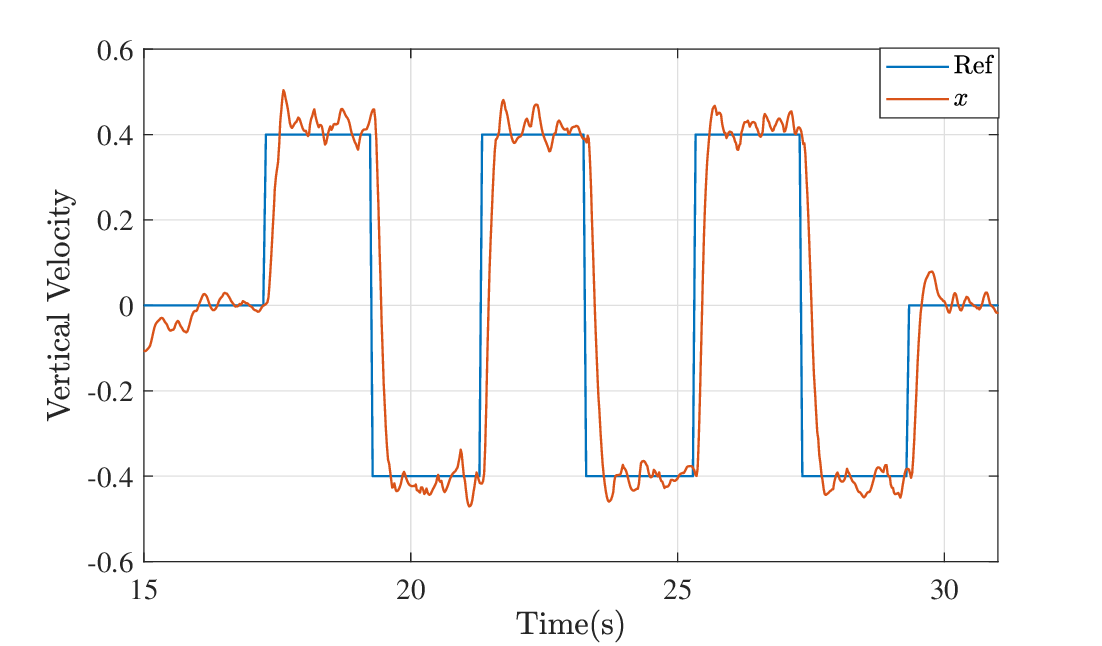}
        \caption{} 
        \label{pid_baseline}
    \end{subfigure}
    \begin{subfigure}[b]{0.49\linewidth}
        \centering
        \includegraphics[width = \linewidth]{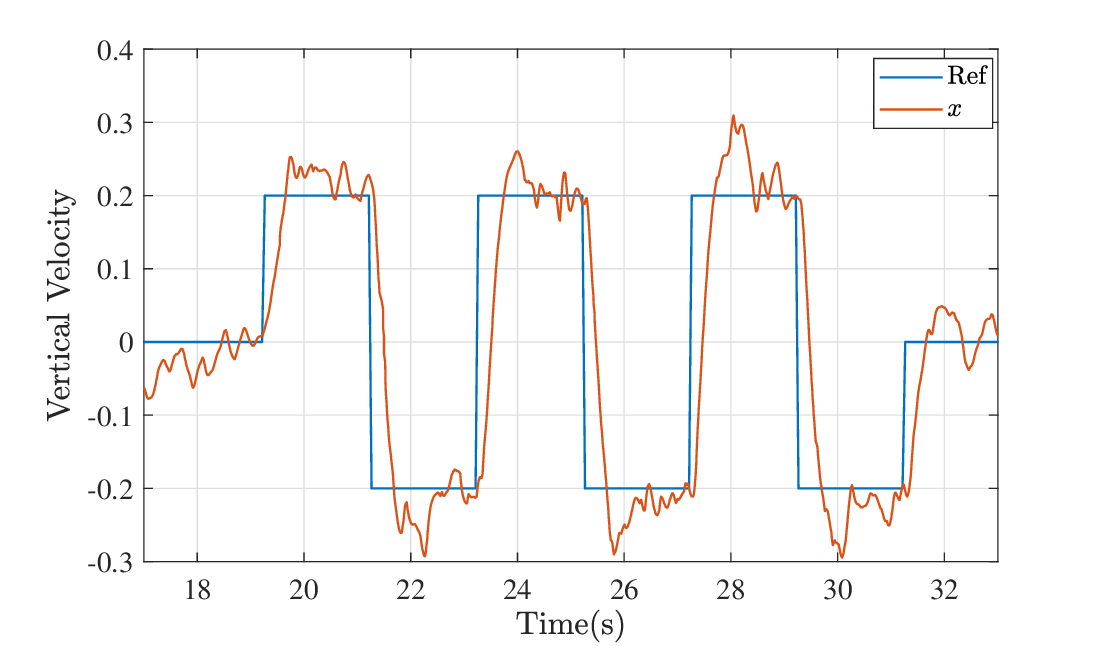}
        \caption{} 
        \label{mrac_baseline}
    \end{subfigure}
    \caption{Vertical velocity $v_z$ [m/s] vs. Time [s]. Reference tracking of baseline controllers: (a) PID, (b) MRAC.} 
    \label{fig:baseline}
\end{figure}

\textit{Experiment 1: Baseline Controllers.} This experiment confirms the controllers' implementation for a homogeneous leader and follower. 
The temporal responses of both PID and MRAC controllers, ~\eqref{eq:PID} and ~\eqref{eq6a}, are exhibited in Figs.~\ref{pid_baseline} and ~\ref{mrac_baseline}, respectively: red lines are the designed controllers, and blue lines are the reference signals. Even though both controllers follow the reference signals well, more elevated overshoots and more delays are observed in the MRAC than in the PID controller. 
The controller~\eqref{eq:PID} has an average overshoot of $25.9\%$ with an average stabilization time of $0.53s$. Conversely, the DMRAC controller~\eqref{eq6a} has an average overshoot value of $26.4\%$ and stabilization time of $0.72s$. The difference between controller performances is mainly due to the processing time required by the DMRAC to perform the mathematical operations associated with the parameters adaptation laws in~\eqref{k1s}, as opposed to the PID, which uses controller parameters calculated offline.  

\begin{figure}[t]
\centering
\includegraphics[width = 0.9\linewidth]{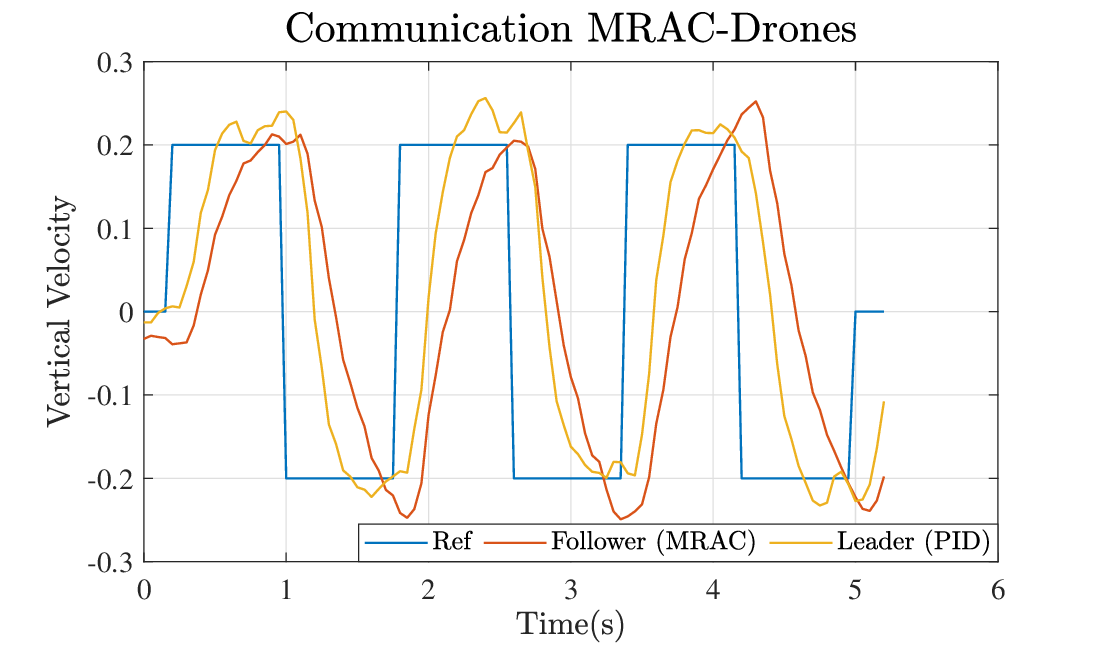}
\caption{Leader-follower communication.}	
\label{fig:communication}
\end{figure}

\textit{Experiment 2: Communication.} We investigate the communication between leader and follower drones conducted by a computer system that acts as a central drone communication server. Based on the knowledge that the controller ~\eqref{eq:PID} performs better than the controller ~\eqref{eq6a} in terms of the settling time and overshoots in \textit{Experiment 1}, the controllers, ~\eqref{eq:PID} and ~\eqref{eq6a}, were applied to the leader and follower drone, respectively.
Fig.~\ref{fig:communication} illustrates the time evolution of the vertical velocity for the leader (yellow) and the follower (red) drone, respectively, with a square wave as a reference signal (blue). Synchronization is achieved in this case: a slight delay between the leader's and follower's response, around $0.1s$ in this experiment, is typical and not considered high in quadcopter applications~\cite{huang2019leader}. Additionally, the leader drone with the controller~\eqref{eq:PID} tracks the reference trajectory with an overshoot of $13.94\%$ and stabilization time of $0.55s$. Likewise, for the follower drone with the controller~\eqref{eq6a}, the average overshoot is $6.37\%$, and a stabilization time of $0.8s$.

\begin{figure}
    \centering
    \begin{subfigure}[b]{0.9\linewidth}
        \centering
        \includegraphics[width = \linewidth]{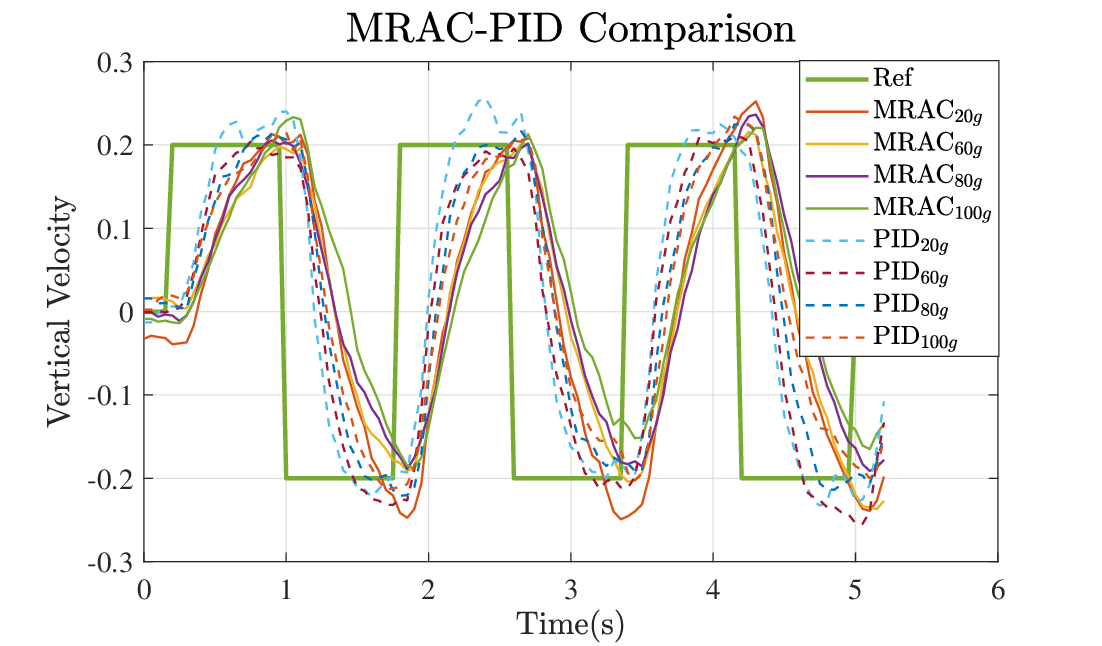}
        \caption{}
        \label{fig:comparison}
    \end{subfigure}\\
    \begin{subfigure}[b]{0.9\linewidth}
        \centering
        \includegraphics[width = \linewidth]{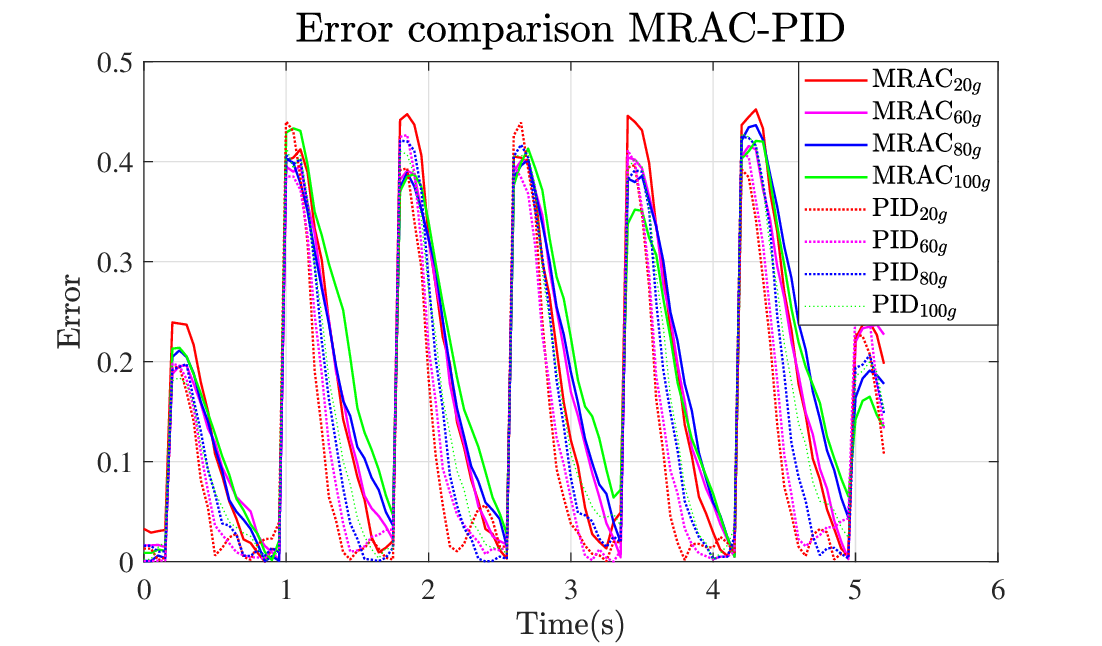}
        \caption{} 
        \label{error_comparison}
    \end{subfigure}
    \caption{MRAC-PID Comparison: (a) Vertical velocity [m/s]; (b) error measure [m/s] vs.Time [s].} 
\end{figure}

\begin{figure}[t]
\centering
\includegraphics[width = 0.9\linewidth]{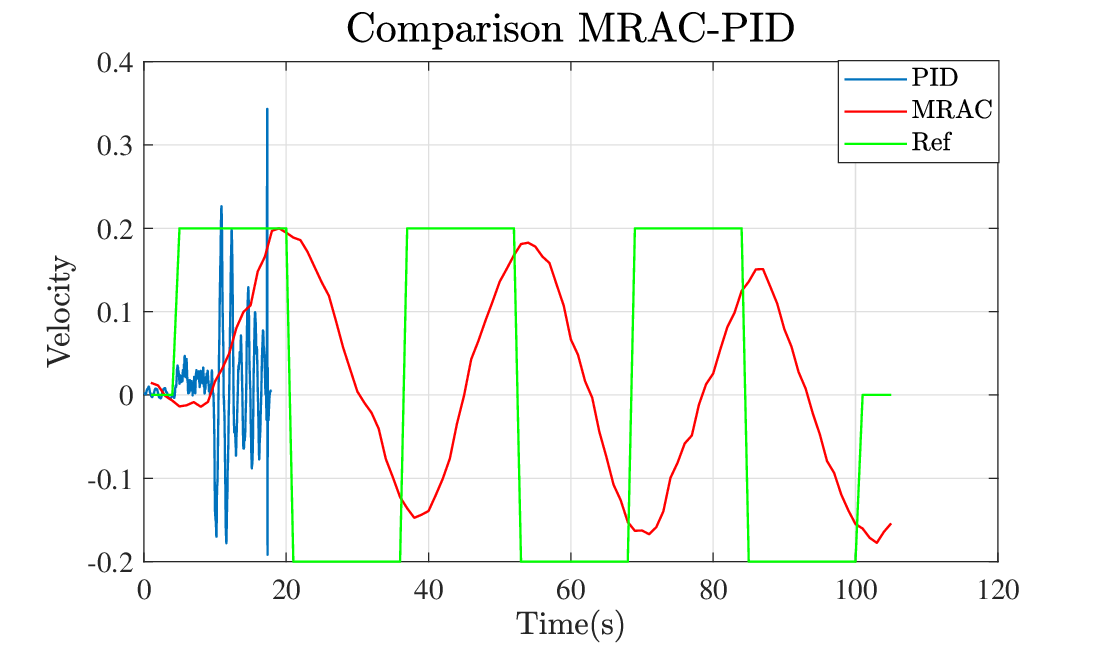}
\caption{Response to a square wave, PID-MRAC comparison. Velocity [m/s] vs. Time [s].}	
\label{fig:failure}
\end{figure}

\textit{Experiment 3: Comparison MRAC-PID.} An experiment is carried out to compare the performance of the adaptive and classical controllers in response to physical parameter changes. We considered changing the mass properties of the drones by gradually adding more weight to the platforms until either drone failed to track the reference signals. The masses added were $20g$, $60g$, $80g$, $100g$, and $120g$, respectively. The temporal response of this set of experiments is shown in Fig.~\ref{fig:comparison}. The solid lines describe the response of controller~\eqref{eq:PID}, while the dotted lines describe the response of controller~\eqref{eq6a} with the different colors for each weight set. 
The average overshoot of the adaptive controller is $6.85\%$, and the average stabilization time is $0.71s$. The average overshoot for the classical PID controller is $12.13\%$, and the stabilization time is $0.64s$. These average values indicate that the overshoot of the adaptive controller is lower than the classical controller; however, its establishment time is a little higher. The synchronization error likewise is computed in Fig.~\ref{error_comparison}.
Comparing the errors does not provide an explicit result of one controller performing better; the difference between both controllers' errors falls on the robustness of the PID controllers~\cite{vilanova2012robustness}. Furthermore, Fig.~\ref{fig:failure} demonstrates the failure of the conventional PID controller~\eqref{eq:PID} with a maximum weight of $120g$ to follow the reference input accurately. In contrast, the MRAC~\eqref{eq6a} successfully tracks the input, and its adaptability is ascertained.

\begin{figure}[t]
\centering
\includegraphics[width = 0.9\linewidth]{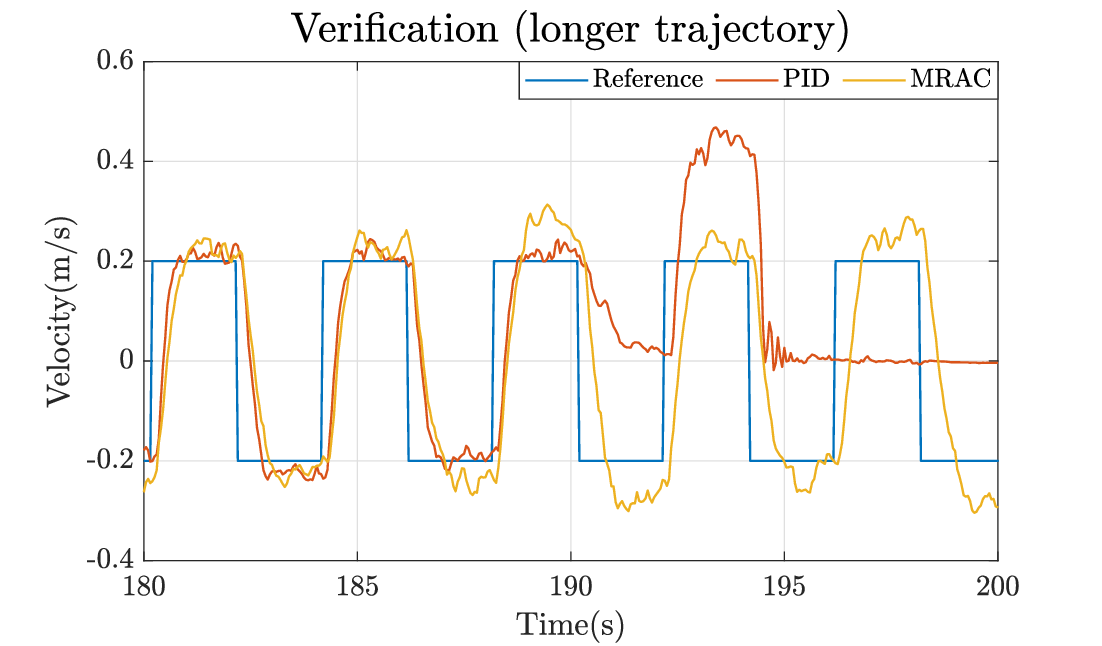}
\caption{Battery life experiment PID-MRAC comparison.}	
\label{fig:battery-life}
\end{figure}

\begin{figure}[t]
\centering
\includegraphics[width = 0.9\linewidth]{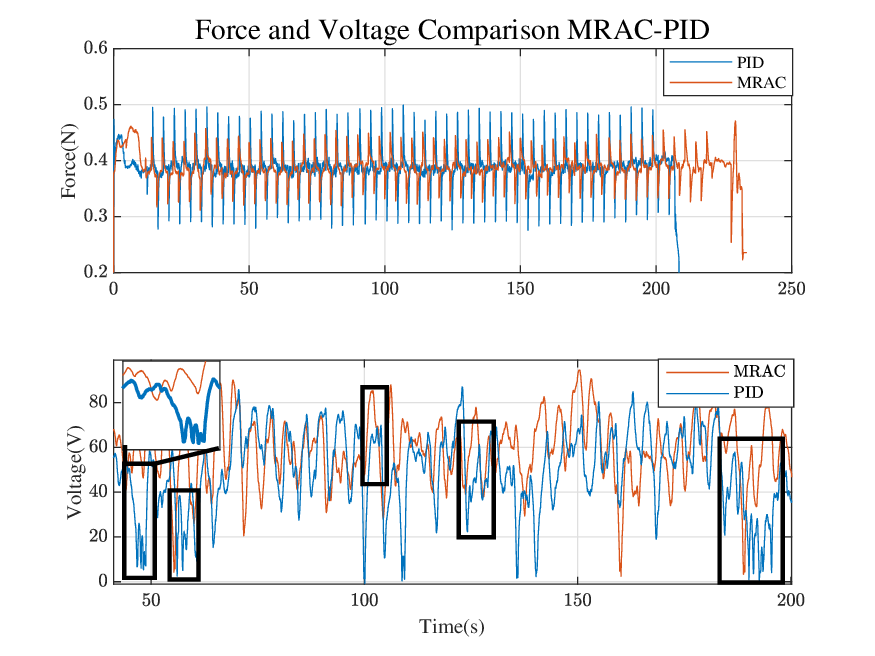}
\caption{Force $F_z$ [N] (top) and Voltage [V] (bottom) to the motors commanded by the PID (blue) and MRAC (red).}	
\label{fig:forceandvoltage}
\end{figure}

\textit{Experiment 4: Power consumption experiment.} We investigate power consumption to discover another factor for a more decisive determination in comparing classic and adaptive controllers. The square wave input signals are applied until either controller's battery is fully depleted in this experiment. In Fig.~\ref{fig:battery-life}, which displays the time-dependent response of the experiment, the controller~\eqref{eq:PID} fails around $190s$, whereas the controller~\eqref{eq6a} still tracks the reference input signal. Based on information on the controller's failure, the controllers' power consumption is compared by exploring the control action, normalized forces from 0 to 1 acting on the drone, and the total applied voltage to four motors directly related to the power(W). The force plot displays the data during the whole period of the experiment, while the voltage plot displays a period of the experiment for a clear view. We observe a force amplitude between $0.3$ and $0.5$ for controller~\eqref{eq:PID}, and around $0.4$ for controller~\eqref{eq6a} in Fig.~\ref{fig:forceandvoltage}. The variance is estimated to quantify the dispersion in each case's control action: $0.0020$ for controller~\eqref{eq:PID} and $0.0012$ for controller~\eqref{eq6a}. Additionally, the larger power consumption in the controller~\eqref{eq:PID} is confirmed in the voltage, which has a proportional relationship to the power (W), through analysis of voltage fluctuations with the controller~\eqref{eq:PID} in Fig.~\ref{fig:forceandvoltage}. As stated in ~\cite{VoltageFluc}, voltage fluctuations refer to unpredictable variations in voltage resulting from many circumstances, such as sudden changes in load. These fluctuations can lead to equipment malfunction and reduce the voltage and current stability within the equipment. In the voltage plot, the black boxes display significant voltage fluctuations, and the first black box is zoomed out to show more voltage changes in PID than MRAC. Regarding the observations of applied forces and voltages to the motors, the adaptive technique consumes less power than a classic controller; controller~\eqref{eq6a} is more efficient than the controller~\eqref{eq:PID}.

\begin{figure}[t]
\centering
\includegraphics[width = 0.9\linewidth]{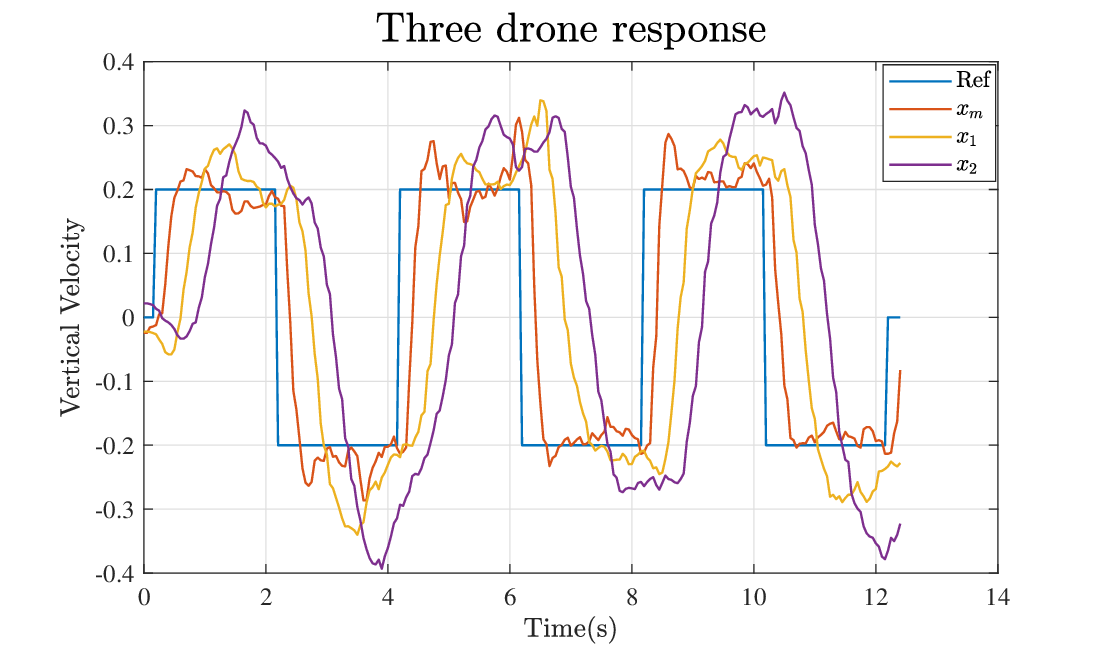}
\caption{Experiment with drones not communicating with the leader. Vertical velocity [m/s] vs. Time [s].}	
\label{fig:threedrones}
\end{figure}

\textit{Experiment 5: Three drone network configuration.} The distributed controller~\eqref{eq10} is verified through physical experiments with three drones. In this experiment, the reference drone transmits its state information and control actions to the first-follower drone, and the first-follower drone transmits its state information and control actions to the second-follower drone; the second-follower drone does not communicate directly with the reference drone. Fig.~\ref{fig:threedrones} displays the synchronization of three. Even though the followers show delays from the one who sends the information to each of them, as explained in \textit{Experiment 2}, the second follower presented an average overshoot of $35.57\%$ with an average stabilization time of $2.12s$. Compared to the previous experiments, the propagation of the error when no direct communication affects the performance but also allows synchronization.

The summary of these experiments at the level of overshoot and establishment time is seen in TABLE~\ref{table_1}. Likewise, the videos of the operation of some of these experiments are found in the following URL:\textit{https://tinyurl.com/Multi-Agent}.

\begin{table}[htbp]
\centering
\begin{tabular}{c | c | c | c | c}
\# & PID Overshoot & MRAC Overshoot & PID $T_s$ & MRAC $T_s$   \\
\hline
1 & 25.9 & 26.4 & 0.53 & 0.72 \\
2 & 13.94 & 6.37 & 0.55 & 0.8 \\
3 & 12.13 & 6.85 & 0.64 & 0.71 \\
\end{tabular}
\caption{Performance comparison table. It is important to note that the values shown are averages over the follower drone.}
\label{table_1}
\end{table}

    \section{Conclusions}\label{s:conclusions}

This work introduces an experimental framework for a distributed model reference adaptive control (DMRAC) strategy for multi-agent synchronization in drone networks. The proposed approach addresses the challenges of real-life parameter variations among individual drones, enhancing the overall system's performance. Implementing a distributed adaptive strategy equips each drone with a model reference adaptive controller that continuously updates its parameters, ensuring adaptability to real-time feedback and reference model information. This adaptability proves vital for maintaining robust performance in the face of uncertainties and discrepancies in the physical parameters of individual drones. The experimental results, particularly in vertical velocity control, validate the effectiveness of the proposed approach in achieving synchronized behavior among the drones, highlighting its potential for practical application in diverse scenarios with inherent differences in drone dynamics. Experiments confirm the proposed approach achieves superior energy management compared to conventional PID techniques, resulting in extended task durations for the drone network. 

Future research should focus on incorporating robust parameters or exploring applications in time-varying networks. Additionally, practical considerations of optimal and predictive control approaches should be considered.

    \bibliographystyle{ieeetr}
\bibliography{main}

\end{document}